%% file: main.tex
\newif\ifiacr
\newif\ifarxiv
\newif\ifacm

\arxivtrue
\acmfalse
\iacrfalse

\ifiacr
\documentclass{llncs}
\usepackage{caption}
\usepackage{subcaption}
\usepackage{cite}
\else

\documentclass[sigconf]{acmart}
\usepackage{balance}

\usepackage{caption, subcaption}
\fi

\def\BibTeX{{\rm B\kern-.05em{\sc i\kern-.025em b}\kern-.08emT\kern-.1667em\lower.7ex\hbox{E}\kern-.125emX}}

\usepackage{microtype}
\usepackage{graphicx}
\usepackage{tablefootnote}
\usepackage{makecell}
\usepackage{booktabs} 
\usepackage{multirow}
\usepackage{listings}
\usepackage{graphicx}
\usepackage{color}
\usepackage{algorithm}
\usepackage{algpseudocode}
\usepackage{hyperref}

\usepackage{pifont}
\newcommand{\cmark}{\ding{51}} 
\newcommand{\xmark}{\ding{55}} 
\usepackage{stfloats} 
\usepackage{soul} 
\usepackage{tikz}
\usepackage[normalem]{ulem} 

\usepackage[separate-uncertainty=true,load=abbr.,load=prefixed,decimalsymbol=fullstop,digitsep=comma,sepfour,per-mode=fraction,binary-units=true]{siunitx}
\usepackage{siunitx}
\usepackage{amsmath}
\usepackage{amsfonts}
\usepackage{amssymb}
\usepackage[nointegrals]{wasysym} 
\usepackage{booktabs}

\newunit[unitsep=space]{\imagespersec}{images\per\second}
\newunit[unitsep=space]{\impersec}{im\per\second}
\newunit[unitsep=space]{\msperimage}{\ms\per image}
\newunit[unitsep=space]{\gbperimage}{\giga\byte\per image}
\newunit[unitsep=space]{\mbperimage}{\mega\byte\per image}
\newunit[unitsep=space]{\mbperim}{\mega\byte\per im}

\newcommand{\Dec}{\texttt{Dec}}
\newcommand{\ct}{\texttt{ct}}
\newcommand{\pt}{\texttt{pt}}
\newcommand{\CC}{\mathbb{C}}
\newcommand{\psp}{p_{sp}}
\newcommand{\ZZ}{\mathbb{Z}}
\newcommand{\Round}[1]{\Bigl\lfloor #1 \Bigr\rceil}

\newcommand{\tocheck}[1]{#1}

\ifiacr
\else
\newtheorem{lemma}{Lemma}
\fi

\definecolor{mygreen}{rgb}{0,0.6,0}
\definecolor{mygray}{rgb}{0.5,0.5,0.5}
\definecolor{mymauve}{rgb}{0.58,0,0.82}

\ifarxiv
\renewcommand\footnotetextcopyrightpermission[1]{} 
\fi

\ifiacr
\fi

\ifacm
\copyrightyear{2019} 
\acmYear{2019} 
\acmConference[WAHC'19]{7th Workshop on Encrypted Computing \& Applied Homomorphic Cryptography}{November 11, 2019}{London, United Kingdom}
\acmBooktitle{7th Workshop on Encrypted Computing \& Applied Homomorphic Cryptography (WAHC'19), November 11, 2019, London, United Kingdom}
\acmPrice{15.00}
\acmDOI{10.1145/3338469.3358944}
\acmISBN{978-1-4503-6829-2/19/11}
\fi

\begin{document}

%
\newcommand{\LongTitle}{nGraph-HE2: 
A High-Throughput Framework for Neural Network Inference on Encrypted Data}
\newcommand{\ShortTitle}{nGraph-HE2}

\ifiacr
\else


\title[\ShortTitle]{\LongTitle}

\author{Fabian Boemer}
\email{fabian.boemer@intel.com}
\affiliation{%
\institution{Intel AI Research}
\city{San Diego}
\state{California}
\country{USA}
}

\author{Anamaria Costache}
\email{anamaria.costache@intel.com}
\affiliation{%
\institution{Intel AI Research}
\city{San Diego}
\state{California}
\country{USA}
}

\author{Rosario Cammarota}
\email{rosario.cammarota@intel.com}
\affiliation{%
\institution{Intel AI Research}
\city{San Diego}
\state{California}
\country{USA}
}

\author{Casimir Wierzynski}
\email{casimir.wierzynski@intel.com}
\affiliation{%
	\institution{Intel AI Research}
	\city{San Diego}
	\state{California}
	\country{USA}
}
\fi

\newcommand\blfootnote[1]{%
	\begingroup
	\renewcommand\thefootnote{}\footnote{#1}%
	\addtocounter{footnote}{-1}%
	\endgroup
}

\ifacm
\renewcommand{\shortauthors}{Boemer, et al.}
\fi

\begin{abstract}
\input{abstract}
\ifarxiv
 \blfootnote{To appear in the 7th Workshop on Encrypted Computing \& Applied Homomorphic Cryptography (WAHC 2019).}
 \fi
\end{abstract}

\ifarxiv

\ccsdesc[500]{Mathematics of computing~Mathematical software}
\ccsdesc[500]{Security and privacy~Privacy-preserving protocols}
\fi

\keywords{Privacy-Preserving Machine Learning; Deep Learning; Graph Compilers; Homomorphic Encryption}

%
\ifiacr
\else
\maketitle
\fi

\input{body}


\clearpage
\ifiacr
\else
\bibliographystyle{ACM-Reference-Format}
\fi
\bibliography{ref}

\appendix
\include{appendix}


\end{document}

%% file: abstract.tex
In previous work, Boemer et al. introduced nGraph-HE, an extension to the Intel
nGraph deep learning (DL) compiler, that enables data scientists to deploy
models with popular frameworks such as TensorFlow and PyTorch with minimal code
changes. However, the class of supported models was limited to relatively
shallow networks with polynomial activations. Here, we introduce \ShortTitle,
which extends nGraph-HE to enable privacy-preserving inference on standard,
pre-trained models using their native activation functions and number fields
(typically real numbers). The proposed framework leverages the CKKS scheme,
whose support for real numbers is friendly to data science, and a client-aided
model using a two-party approach to compute activation functions.

We first present CKKS-specific optimizations, enabling a 3x-88x runtime speedup
for scalar encoding, and doubling the throughput through a novel use of CKKS
plaintext packing into complex numbers. Second, we optimize ciphertext-plaintext
addition and multiplication, yielding 2.6x-4.2x runtime speedup. Third, we
exploit two graph-level optimizations: \emph{lazy rescaling} and
\emph{depth-aware encoding}, which allow us to significantly improve performance.

Together, these optimizations enable state-of-the-art throughput of
{\SI[allow-number-unit-breaks]{1998}\imagespersec} on the CryptoNets network. Using the client-aided model, we also present
homomorphic evaluation of (to our knowledge) the largest network to date,
namely, pre-trained MobileNetV2 models on the ImageNet dataset, with
60.4\percent/82.7\percent\ top-1/top-5 accuracy and an amortized runtime of
\SI{381}\msperimage.

%% file: body.tex
\input{introduction}

\input{background}

\input{implementation}

\input{results}

\input{extensions}

%
%

%% file: introduction.tex
\section{Introduction}
The proliferation of machine learning inference as a service raises privacy
questions and concerns. For example, a data owner may be concerned
about allowing an external party access to her data. 


Homomorphic encryption (HE) is an elegant cryptographic technology which can
solve the data owner's concern about data exposure. HE is a form of encryption
with the ability to perform computation on encrypted data, without ever
decrypting it. In particular, HE allows for a data owner to encrypt her data,
send it to the model owner to perform inference, and then receive the encrypted
inference result. The data owner accesses the result of the inference by
decrypting the response from the server.

The class of HE schemes known as leveled HE schemes or somewhat HE
(SHE) schemes supports a limited number of additions and multiplications. As such, these
schemes are attractive solutions to the DL based inference, whose core workload
is multiplications and additions in the form of convolutions and generalized
matrix multiplications (GEMM). One challenge in enabling HE for DL using SHE
schemes is that we cannot compute non-linear functions, common in deep neural
networks activations.

Another challenge in enabling HE for DL is the lack of support in existing
frameworks. While popular DL frameworks such as
TensorFlow~\cite{abadi2016tensorflow} and PyTorch~\cite{pytorch} have greatly
simplified the development of novel DL methods, they do not support
HE. Meanwhile, existing
HE libraries such as Microsoft SEAL~\cite{sealcrypto}, HElib~\cite{halevi2018faster}, and
Palisade~\cite{palisade} are typically written at a level far lower
than the primitive operations of DL. As a result, implementing
DL models in HE libraries requires a significant engineering overhead.

nGraph-HE~\cite{boemer2019ngraph} introduced the first industry-class,
open-source DL graph compiler which supports the execution of DL models through
popular frameworks such as TensorFlow, MXNet, and PyTorch. Graph compilers
represent DL models using a graph-based intermediate representation (IR), upon
which hardware-dependent and hardware-agnostic graph optimizations are
performed. By treating HE as a virtual hardware target, nGraph-HE takes
advantage of the graph compiler toolchain to create a framework for DL with HE.
nGraph-HE uses Microsoft SEAL~\cite{sealcrypto} for the underlying HE evaluation 
(with a framework for additional HE schemes), and
nGraph~\cite{cyphers2018intel} for the graph compiler IR.
nGraph-HE enabled data scientists to use familiar DL frameworks; however,
nGraph-HE supported only a limited class of models, restricted to polynomial
activations.

In this work, we present \ShortTitle\footnote{\ShortTitle\ is
	available under the Apache 2.0 license at \url{https://ngra.ph/he}.}, which introduces
a number of optimizations in the graph compiler and the HE library. \ShortTitle\
utilizes a client-aided model, i.e. a hybrid approach using two-party
computation, to execute a much wider class of pre-trained deep neural networks
including non-polynomial activations with a focus on maximizing throughput. Our
optimizations focus on inference on encrypted data with a plaintext model. We
use batch-axis packing (Section~\ref{sec:background_difficulties}) to enable a
simple implementation of the Reshape operation and significantly increase
throughput.

This setting is friendly to data scientists. It supports standard DL
models,including non-polynomial activations. Since we do not rely on HE-specific
training models, the training phase is HE-independent. Thus, data scientists can
perform HE inference on standard DL models without cryptographic expertise.

A challenge specific to this data-scientist-friendly setting is that
neural networks typically contain operations not suitable to all HE schemes,
particularly in the activation functions. For instance, computing ReLU or
MaxPool requires the comparison operator, which is not supported in the CKKS HE
scheme. To this end, we use a protocol in which the server interacts with a
client to perform non-polynomial operations such as ReLU
(Section~\ref{sec:client_server}). Nevertheless, the CKKS scheme has several
advantages, including support for floating-point numbers, plaintext packing, and
faster runtime.

We present three main contributions. First, we describe optimizations to the
CKKS encoding operations in SEAL (Section~\ref{sec:ckks_encoding_opt}). We
demonstrate a 3x-88x improvement in scalar encoding, and introduce \emph{complex
	packing}, an optimization which doubles the inference throughput in networks
without ciphertext-ciphertext multiplication
(Section~\ref{sec:complex_packing}). Second, we introduce optimizations to
ciphertext-plaintext addition, and ciphertext-plaintext multiplication, which
apply in the batch-axis plaintext packing setting
(Section~\ref{sec:ckks_arithmetic_optimizations}). Third, we exploit two
graph-level optimizations (Section~\ref{sec:graph_opt}). The first graph-level
optimization, \emph{lazy rescaling}, improves the runtime of higher-level
operations such as Dot and Convolution\footnote{Dot is a generalized dot
	product operation and Convolution is a batched convolution operation; see
	\url{https://www.ngraph.ai/documentation/ops} for more
	information.} by delaying, hence minimizing the runtime spent on, the expensive
rescaling operation. The second graph-level optimization, \emph{depth-aware
	encoding}, minimizes the memory usage of the encoded model by encoding at the
appropriate coefficient modulus level. Our \emph{just-in-time encoding}
implementation of depth-aware encoding encodes the values as late as possible.

We evaluate our contributions on both small, single-operation tests
(Section~\ref{sec:result_low_level_operations}), and on larger neural networks
(Section~\ref{sec:nn_evaluation}). In particular, we demonstrate
state-of-the-art performance on the CryptoNets network
(Section~\ref{sec:cryptonets_evaluation}), with a throughput of
{\SI[allow-number-unit-breaks]{1998}\imagespersec}. Our contributions also
enable the first, to our knowledge, homomorphic evaluation of a network on the
ImageNet dataset, MobileNetV2, with 60.4\percent/82.7\percent\ top-1/top-5
accuracy and amortized runtime of \SI{381}\msperimage\
(Section~\ref{sec:mobiletnetv2}). This is the first work showing the
privacy-preserving execution of a full production-level deep neural network.

%% file: background.tex
\section{Background}

\subsection{Homomorphic Encryption}
Homomorphic encryption (HE) enables computations to be carried out on
encrypted data. We will focus on the FV scheme (sometimes referred to as BFV)
~\cite{brakerski2012fully, fan2012somewhat}, as implemented in SEAL version 3.3
~\cite{sealcrypto}, with the CKKS optimizations ~\cite{cheon2017homomorphic,
	cheon2018full} and the relinearization technique with the special
prime~\cite{chenefficient}. This is a somewhat HE (SHE) scheme, meaning that it
supports a limited (and pre-determined) number of additions and multiplications.
In contrast, fully homomorphic encryption (FHE) schemes support an unlimited
number of additions and multiplications, typically by modifying an SHE scheme 
with an expensive bootstrapping step. 

More concretely, if $\ct_1$ and $\ct_2$ are encryptions of
$m_1$ and $m_2$, respectively, then
\begin{equation}
\label{eq:he_cc_add_mult}
 \Dec(\ct_1 + \ct_2) \approx m_1 + m_2,  \quad \Dec(\ct_1 \cdot \ct_2) \approx m_1 \cdot m_2       
 \end{equation}
The imprecision in the arithmetic is due to noise introduced during the computation,
and can be controlled by setting encryption parameters appropriately. CKKS also offers ciphertext-plaintext operations, which are typically faster than ciphertext-ciphertext operations in Equation~\ref{eq:he_cc_add_mult}. That is, if $\pt_1$ is an encoding of $m_1$, then
\[     \Dec(\pt_1 + \ct_2) \approx m_1 + m_2, \quad  \Dec(\pt_1 \cdot \ct_2) \approx m_1 \cdot m_2   \]
\subsection{Mathematical Background}
\label{sec:math_background}
Many homomorphic encryption schemes, including CKKS, are based on the ring learning
with error (RLWE) problem. A full description of the CKKS scheme and the RLWE problem 
is outside the scope of this paper. Instead, we provide a brief introduction to the 
CKKS scheme and refer the reader to~\cite{cheon2017homomorphic, LPR10} for additional
details.

Let $\Phi_M(X)$ be the M$^{\text{th}}$ cyclotomic polynomial of degree $N =
\phi(M)$. The plaintext space is the ring $\mathcal{R} = \ZZ [X]/ (\Phi_M(X))$. We 
always take to be $\deg(\Phi_M(X))$ a power of two, typically $\SI{2048}{}$ or $\SI{4096}{}$. This
is for both performance and security reasons.

\subsubsection{Rescaling}
\label{sec:rescaling}
In most HE schemes, a message is encrypted by adding noise. This
noise grows with each homomorphic operation, especially multiplication. To
manage this noise growth, CKKS introduces a rescaling operation which lowers the
noise, and is typically performed after every multiplication.

We can only perform a (predetermined) limited number of such rescaling
operations; therefore we can perform a (predetermined) number of
multiplications. We let $L$ be this number. Each multiplication represents a
`level' in our ciphertext space, and a rescaling operation lowers the level. To
implement this, we have a `layered' ciphertext space, where each layer has a
different ciphertext modulus. We construct this space as follows. Let $p_1, \ldots,
p_L$ be primes, and let $\psp$ be a `special prime.' The ciphertext modulus is
$q_L = \prod_{i=1}^L p_i$, yielding ciphertext space $\mathcal{R}_{q_L} =
\mathcal{R} / (q_L\mathcal{R})$. Ciphertexts in the CKKS scheme are typically
pairs of polynomials, i.e., $\ct \in \mathcal{R}_{q_L}^2$. The relinearization
step (also referred to as the key-switching step) is performed
using the raise-the-modulus idea from~\cite{GHS12} and the special modulus
$\psp$.

Encryption is performed using the special prime; this means a fresh ciphertext will be modulo $q_L\cdot\psp$.
We immediately perform a scale operation to reduce the level to that of $q_L$, so that the
encryption algorithm's final output is an element of $\mathcal{R}_{q_L}$.

The rescaling algorithm is the homomorphic equivalent to the removing inaccurate
LSBs as a rounding step in approximate arithmetic. More formally, we bring a
ciphertext $\ct$ from level $\ell$ to $\ell'$ by computing

\begin{equation}
\label{eq:rescaling}
  \ct' \leftarrow \Round{\ct\frac{q_\ell}{q_\ell'}}  
\end{equation}
where $q_{\ell} = \prod_{i=1}^{\ell} p_i$. Typically, rescaling is performed with $\ell' = \ell-1$ after each multiplication to minimize noise growth. As such, the encryption parameter $L$
is typically set to be at least $L_f$, i.e. $L \geq L_f$, the multiplicative depth of the function to
compute.

\subsubsection{Double-CRT Representation}
\label{double_crt}
To enable fast modular arithmetic modulo large integers, SEAL uses the residue 
number system (RNS) to represent the integers.
To use this, we choose the factors $q_i$ in $q_{\ell} = \prod_{i=1}^\ell p_i$ to
be pairwise coprime, roughly of the same size and 64-bit unsigned integers
(they are typically chosen to be of size 30-60 bits). Then, using the Chinese
remainder theorem, we can write an element $x$ in its RNS representation (also referred to as the CRT
	representation.) $(x \pmod{q_i})_i$. Each operation on $x$ can be implemented by
applying the operation on each element $x_i$. In particular, addition and
multiplication of two numbers in RNS form are performed element-wise in $O(L)$
time, rather than $O(L \log L)$ time for multiplication, as would be required in
a naive representation.

SEAL also implements the number-theoretic transform (NTT) for fast polynomial
multiplication. Together, the CRT and NTT representation is known as the
`double-CRT' form. However, the NTT representation is incompatible with the rescaling
operation. SEAL's rescaling operation requires performing an NTT$^{-1}$, the
rescaling operation (\ref{eq:rescaling}), then an NTT. The NTT and its inverse are relatively expensive computations, hence we will describe optimizations for avoiding them where possible (Section~\ref{sec:graph_opt}). A full
description of the NTT is beyond the scope of this paper; see for
example~\cite{LongaNaehringNTT} for a cryptographer's perspective of the NTT.

 \subsubsection{Plaintext Packing} 
 \label{sec:plaintext_packing}
An extremely useful feature of the CKKS scheme is plaintext packing,
also referred to as batching. This allows us to “pack” $N/2$ complex scalar
values into one plaintext or ciphertext, where $N$ is the cyclotomic polynomial
degree. It works by defining an encoding map $ \CC^{N/2} \rightarrow  
\mathcal{R}$, where $\mathcal{R}$ is the plaintext space. An operation
(addition or multiplication) performed on an element in $\mathcal{R}$
corresponds to the same operation performed on $N/2$ elements in $\CC^{N/2}$.
The number $N/2$ elements in the packing is also known as the number of \emph{slots} in the plaintext.

Let $\mathcal{P} = \mathcal{R}$ refer to the plaintext space, and
$\mathcal{C} = \mathcal{R}_{q_L}^*$ refer to the ciphertext space.

\subsection{HE for Deep Learning}
\label{sec:he_for_deep_learning}
The ability of HE to perform addition and multiplication makes it attractive to
DL, whose core workloads are multiplication and addition in the
form of convolutions and GEMM operations. However,
neural networks commonly contain operations not suitable to all HE schemes,
particularly in the activation functions. For instance, computing ReLU or
MaxPool requires the comparison operator, which is not supported in all SHE
schemes. At a high level, therefore, there are two broad approaches to enabling 
homomorphic evaluation of a given DL model:
\begin{enumerate}
	\item {\it HE-friendly networks}: Modify the network to be HE-friendly, and re-train.
	\item {\it Pre-trained networks}: Modify the HE scheme or protocol to accommodate the network as is, ideally with no retraining. 
\end{enumerate}

\subsubsection{HE-friendly Networks}
In this setting, we assume (and require) that the data scientist has access to the
entire DL workflow, including training. Here, the network is re-trained with
polynomial activations, and max-pooling is typically replace with average
pooling. Low-degree polynomials may be used, as high-degree polynomials result
in prohibitively large encryption parameters due to the large multiplicative depth.
The CryptoNets network~\cite{gilad2016cryptonets} is the seminal HE-friendly
network, using the $f(x) = x^2$ activation function to achieve ${\approx}99\percent\ $
accuracy on the MNIST~\cite{lecun1998mnist} handwritten digits dataset. However,
on larger datasets, the accuracy of HE-friendly networks suffers.
CHET~\cite{dathathri2019chet} adopts a similar approach on the
CIFAR10~\cite{krizhevsky2014cifar} dataset, instead using activation functions
$f(x) = ax^2 + bx$, with $a,b, \in \mathbb{R}$. This approach results in 81.5\percent\
accuracy, down from 84\percent\ accuracy in the original model with ReLU activations.
Hesamifard et al.~\cite{cryptodl_updated} see a similar drop-off in accuracy from
94.2\% to 91.5\% on the CIFAR10 dataset. Depending on the use case, such a drop
in accuracy may not be acceptable.

From a practical viewpoint, HE-friendly networks tend to be more difficult to
train than their native counterparts. In particular, polynomial activations are
unbounded and grow more quickly than standard activation functions such as ReLU
or sigmoid, resulting in numerical overflow during training. Possible
workarounds include weight and activation initialization and gradient clipping.

Sparsification methods, such as in SEALion~\cite{van2019sealion} and Faster
CryptoNets~\cite{chou2018faster} improve latency by reducing the number of
homomorphic additions or multiplications. This is an optimization mostly
independent of HE.

\subsubsection{Pre-trained Networks}
\label{sec:fixed_network}
In this setting, we assume a network has been trained, and no modifications are
possible. This setting results in independent training and inference tasks. In
particular, data scientists need not be familiar with HE to train
privacy-preserving models. Additionally, this setting preserves the accuracy of
the existing models, which tend to be higher than models built with HE-friendly
constraints. Two solutions to the pre-trained network setting are FHE schemes and
hybrid schemes.

\emph{FHE schemes.} FHE schemes enable an unlimited number of additions and
multiplications, allowing for arbitrary-precision polynomial approximations of
non-polynomial activations. However, due to the expensive bootstrapping step
used to cope with increasing the computational depth, this approach is typically
much slower than alternatives. Some FHE schemes, such as TFHE~\cite{tfhe}
operate on boolean circuits which support low-depth circuits for exact
computation of ReLU. However, performance on arithmetic circuits, such as GEMM
operations, suffers.

Wang, et al.~\cite{wang2019toward} propose using Intel Software Guard Extensions
(SGX) to implement a bootstrapping procedure in a trusted execution environment
(TEE). In effect, this approach turns a SHE scheme to a FHE scheme with a lower
performance penalty than FHE bootstrapping. However, it loosens the security
model, as the TEE must be trusted.

\emph{Hybrid schemes.} Hybrid schemes combine privacy-preserving primitives,
such as HE and multi-party computation (MPC). In MPC, several parties follow a
communication protocol to jointly perform the computation. MPC techniques, such
as garbled circuits (GCs), typically support a broader range of operations than
HE, while introducing a communication cost between the
parties. Hybrid HE-MPC schemes therefore provide an elegant solution to the
pre-trained network setting by using MPC to perform non-polynomial activations, and
HE to perform the FC and Convolution layers.

This approach has two important benefits. First, it enables exact computation,
mitigating the performance drop-off in HE-friendly networks. Second, it enables
smaller encryption parameters. The HE-MPC interface involves refreshing the
ciphertext at each non-polynomial activation, i.e. resetting the noise budget
and coefficient modulus to the highest level $L$. This resetting reduces the
effective multiplicative depth of the computation to the number of
multiplications between non-polynomial activations. As a result, $L$ is quite
small, even for large networks. For instance, $L=3$ suffices for the MobileNetV2
network~\cite{sandler2018mobilenetv2} (Section~\ref{sec:mobiletnetv2}). Smaller
$L$ also enables choice of smaller polynomial modulus degree, which greatly
reduces the runtime (see Appendix~\ref{app:seal_perf_test}) and memory usage.

Several hybrid schemes have been developed. Chimera~\cite{bourachimera} is a
hybrid HE-HE scheme which performs ReLU in TFHE, and affine transformations in
an arithmetic-circuit HE scheme such as FV or CKKS. However, the translation
between TFHE and FV/CKKS is potentially expensive. MiniONN~\cite{minionn} is a
hybrid HE-MPC scheme which uses an additive HE scheme to generate multiplication
triples, which are used in an MPC-based evaluation of the network.
Gazelle~\cite{juvekar2018gazelle} uses HE to perform the polynomial functions
and GCs to perform the non-polynomial activations.

\emph{Other schemes.} A third solution to the pre-trained network setting is
pure MPC schemes. ABY~\cite{demmler2015aby} supports switching between arithmetic, boolean, and
Yao's GCs. ABY3~\cite{mohassel2018aby}
increases the performance of ABY by introducing a third party.
SecureNN~\cite{wagh2019securenn} likewise increases performance at
the cost of a third party. Some two-party MPC schemes also have shortcomings,
such as requiring binarizing the network~\cite{riazi2019xonn}. Our work, in contrast, supports full-precision networks using standard data types.

\subsubsection{Challenges in deploying DL on HE} \label{sec:background_difficulties} \hfill 

\textbf{Software Frameworks}. One difficulty in enabling HE for DL is the lack
of support in existing frameworks. While popular DL frameworks such as
TensorFlow~\cite{abadi2016tensorflow} and PyTorch~\cite{pytorch} have greatly
simplified the development of novel DL methods, they do not support HE. Existing
HE libraries such as Microsoft SEAL~\cite{sealcrypto}, HElib~\cite{halevi2018faster},
and Palisade~\cite{palisade} are typically written at a low level. As such,
implementing DL models requires a significant engineering overhead. nGraph-HE
~\cite{boemer2019ngraph} introduces a DL graph compiler which
supports execution of DL models through popular frameworks such as TensorFlow,
MXNet, and PyTorch.

\textbf{Performance Considerations}. One of the primary shortcomings of HE is
the large computational and memory overhead compared to unencrypted computation,
which can be several orders of magnitude. The choice of encryption parameters,
$N$ and the coefficient moduli $q_i$, has a large impact on this overhead, as
well as the security level (see Appendix~\ref{app:seal_perf_test}). As such,
parameter selection, which remains a largely hand-tuned process, is vital for
performance.

\textbf{Mapping to DL Functions.}
Another difficulty in enabling HE for DL is the mapping from HE operations to DL
operations. While HE addition and multiplication map naturally to plaintext
addition and multiplication, there are various choices for plaintext packing (see
Section~\ref{sec:math_background}). Both
CryptoNets~\cite{gilad2016cryptonets} and nGraph-HE~\cite{boemer2019ngraph} use
plaintext packing along the batch axis (\emph{batch-axis packing}) to store a 4D
tensor of shape $(S, C, H, W)$ (batch size, channels, height, width) as a 3D
tensor of shape $(C, H, W)$, with each ciphertext packing $S$ values. Each model
weight is stored as a plaintext with the same value in each slot (encoded
using scalar encoding, see Section~\ref{sec:scalarencoding}). Since HE addition
and multiplication are performed element-wise on each slot, this enables
inference on up to $S$ data items simultaneously, where the runtime for one
data item is the same as for $S$ data items (for $S \leq N/2$, the slot count). As a result, this use of plaintext packing greatly increases throughput for a given latency.

Other approaches such as Gazelle~\cite{juvekar2018gazelle} and
LoLa~\cite{Brutzkus2019LowLatency} use \emph{inter-axis packing}, a choice of plaintext packing which encrypts multiple scalars from the same inference
data item or weight matrix to the same ciphertext. Inter-axis packing optimizes inference on one data item at a time, with latency scaling linearly
with the batch size. However, DL workloads on inter-axis packing often use
HE rotations, which are relatively expensive (see
Appendix~\ref{app:seal_perf_test}). The optimal packing approach depends on the
workload, and can be determined by graph compilers. \ShortTitle\ uses
batch-axis packing.

\subsection{Graph Compilers}
Graph compilers represent DL models with a graph-based intermediate
representation (IR). The primary advantage to a graph-based IR is the enabling
of graph-based optimizations, which can be either hardware agnostic or hardware
dependent. Intel nGraph~\cite{cyphers2018intel} is a DL graph compiler which
optimizes the inference graph for several hardware targets. A second advantage
to graph-based IR is the ability to represent models from different DL
frameworks in a common IR; thus, the graph-based optimizations are
framework-agnostic. nGraph-HE~\cite{boemer2019ngraph} introduces the first
DL framework for HE. nGraph-HE treats HE as a virtual hardware target and uses Microsoft SEAL~\cite{sealcrypto} for the
underlying HE evaluation, as well as a simple structure for adding other HE libraries. In addition to graph-based optimizations, nGraph-HE
provides run-time based optimizations based on the values of the plaintext
model.

CHET~\cite{dathathri2019chet} is another graph-based compiler for HE. It uses
inter-axis packing to optimize the layout of each tensor, as opposed to using
batch-axis packing for every tensor, as in nGraph-HE.
SEALion~\cite{van2019sealion} uses a graph compiler for automatic
parameter selection, while lacking packing and value-based
runtime optimizations.

%% file: implementation.tex
\section{Contributions}
We introduce the following contributions, which apply in the batch-axis packing setting:
\begin{itemize}
	\item CKKS encoding optimizations;
	\item CKKS arithmetic optimizations;
	\item graph-level optimizations.
\end{itemize}
The CKKS encoding optimizations include faster scalar encoding, and
\emph{complex packing}, which doubles the throughput by taking advantage of the
complex components of the plaintext encoding map. Our arithmetic
optimizations apply to ciphertext-plaintext addition, and ciphertext-plaintext
multiplication. The graph-level optimizations include \emph{lazy rescaling} and \emph{depth-aware encoding},
which reduce the runtime spent rescaling and encoding, respectively.

\subsection{CKKS Encoding Optimizations}
\label{sec:ckks_encoding_opt}

\subsubsection{Scalar Encoding}
\label{sec:scalarencoding}
Plaintext packing enables the encoding of $N/2$ complex scalars into a single
plaintext. For more efficient addition and multiplication, SEAL stores each
plaintext in double-CRT form (performing an NTT on the polynomial, and storing
each coefficient in RNS form with respect to the $p_i$). At the top level, (with
$L$ coefficient moduli), encoding requires $O(LN)$ memory and $O(LN \log N)$
runtime. Algorithm~\ref{alg:encode_plain_general} shows the pseudocode for
general encoding, including the NTT.

\begin{algorithm}
	\caption{General CKKS Encoding}
	\label{alg:encode_plain_general}
	\begin{algorithmic}[1]
		\Function{EncodeVector}{$c \in \CC^{N/2}, q \in
			\mathbb{Z}, s \in \mathbb{R}$}
		\State $p \in \CC^{N}$
		\State $p[0:N/2] \gets c$ \label{line:gen_encode:c}
		\State $p[N/2+1:N] \gets c^*$ \label{line:gen_encode:c_conj}
		\State $p \gets \text{DFT}^{-1}(p \cdot s)$ \label{line:gen_encode:invdft}
		\State $p \gets [p]_q$	\label{line:gen_encode:mod}
		\State $p \gets \text{NegacyclicNTT}(p)$ \label{line:negacyclicntt}
		\EndFunction
	\end{algorithmic}
\end{algorithm}

SEAL additionally provides an optimized encoding algorithm in the setting where
the $N/2$ scalars are the same real-valued number. This setting yields a
simplified DFT$^{-1}$ and NTT, resulting in an implementation requiring $O(LN)$
runtime and memory. Both of SEAL's encoding implementations are general, that is
they allow arbitrary operations on the resulting plaintext.

Here, we optimize for the more restrictive case in which $N/2$ identical
real-valued scalars are stored in the plaintext, for the entire lifetime of the plaintext.
Our use of batch-axis packing (see Section~\ref{sec:complex_packing})
maintains this property on the plaintexts, since they are used only for
plaintext addition and multiplication. Other plaintext packing schemes, such as
inter-axis packing (see Section~\ref{sec:background_difficulties}), however, do
not maintain this property. Thus, scalar encoding applies only in specific
use-cases, including batch-axis packing.

Our optimization takes advantage of the fact that the general CKKS encoding algorithm
of $N/2$ identical real-valued scalars will result in a plaintext with $N$
identical values across the slots. See Appendix~\ref{app:scalar_encoding_proof}
for the proof of this property. So, rather than store $N$ copies of the same
scalar, we modify the plaintext to store just a single copy. This improves the
memory usage and runtime each by a factor of $N$, yielding $O(L)$ runtime and
memory usage. Algorithm~\ref{alg:encode_scalar} shows the pseudocode for the scalar-optimized encoding algorithm.

\begin{algorithm}
	\caption{CKKS Scalar encoding of $c$ with respect to modulus $q$ at scale $s$}
	\label{alg:encode_scalar}
	\begin{algorithmic}[1]
		\Function{EncodeReal}{$c \in \mathbb{R}, q \in \mathbb{Z}, s \in \mathbb{R}$}
		\State $y \in \mathbb{R}$
		\State $y \gets [s\cdot c]_q$
		\State
		\Return $y$
		\EndFunction
	\end{algorithmic}
\end{algorithm}
Note, SEAL implements a variant of Algorithm~\ref{alg:encode_scalar} for
scalar encoding; however it computes $y \in \mathbb{R}^{N}$, with $y_i \gets
[s \cdot c]_q\ \forall i$, requiring memory and runtime $O(LN)$. For comparison,
Algorithm~\ref{alg:encode_scalar} avoids the expensive copy of size $N$,
decreasing the runtime compared to SEAL's implementation.

\subsubsection{Complex packing}
\label{sec:complex_packing}
We introduce \emph{complex packing}, an optimization which doubles the inference
throughput in cases without ciphertext-ciphertext multiplication. One of the
primary ways to combat the large runtime and memory overhead of HE is to use
plaintext packing in the CKKS encoding mapping $\mathbb{C}^{N/2} \rightarrow
\mathcal{R}$ (Section~\ref{sec:plaintext_packing}). Neural network models,
however, typically operate on real numbers. As such, packing real-valued model
weights or data values utilizes at most half of the available computation.
Complex packing, on the other hand, utilizes the entire computational capacity
of plaintext packing.

For simplicity, let $N=4$, so each plaintext and ciphertext encodes two
complex scalars. Given  scalars $a, b, c, d, f, g, h, k \in \mathbb{R}$,  let:
\begin{itemize}
	\item $REnc(a,b) = p(a,b)$ represent the plaintext encoding with $a$ in the
	first slot and $b$ in the second slot.
	\item $CEnc(a,b, c, d) = p(a+bi, c+di)$ encode $a+bi$ in the first slot, and
	$c+di$ in the second slot.
	\item $RDec(p(a,b)) = (a,b)$
	\item $CDec(p(a+bi,c+di)) = (a, b, c, d)$
\end{itemize}
Let \emph{real packing} refer to the $REnc/RDec$ representation, and
\emph{complex packing} refer to the $CEnc/CDec$ representation. Then, let
\[
\begin{array}{l l l l  }
&p(a+bi, &&c+di) \\
&p(f+gi, &&h+ki) \\
\cline{1-4}
\stackrel{\pm}{\rightarrow} &p(a \pm f + (b \pm g)i, &&c \pm h + (d \pm k)i) \\
\stackrel{\times}{\rightarrow} &p(af -bg + (ag+bf)i, && ch-dk + (ck+dh)i)
\end{array}
\]
represent element-wise (real) addition/subtraction and multiplication,
respectively. Note, a given implementation of a plaintext may not represent the
plaintext slots internally as complex numbers. SEAL, for instance, uses 64-bit
unsigned integers. Instead, our presentation serves to illustrate the concept,
which is independent of the HE library's specific implementation.
Though we only consider plaintexts here, the same
logic also holds for ciphertexts.

Now, we consider the following element-wise computations:
\begin{itemize}
	\item Add/Subtract:
	\begin{align*}
	(a, b, c, d) \pm (f, g, h, k) &= (a \pm f, b \pm g, c \pm h, d \pm k) \\
	&= CDec(CEnc(a,b, c, d) \pm CEnc(f, g, h, k))
	\end{align*}
	\item Broadcast-Multiply: \begin{align*}
	(a, b, c, d) \times f &= (af, bf, cf, df) \\
	&= CDec(CEnc(a,b, c, d) \times CEnc(f,0,f,0))
	\end{align*}
	\item Multiply: \begin{align*}
	(a, b, c, d) \times (f, g, h, k) &= (af, bg, ch, dk)) \\
	&\neq CDec(CEnc(a,b,c,d) \times CEnc(f,g,h,k))
	\end{align*}
\end{itemize}
So, we observe each operation except Multiply\footnote{due to the cross-terms in
	$(a+bi)\times (f+gi) = af-bg +(ag+bf)i \neq af+bgi$. Note, it is an open problem to compute and add the correction term $bg + (bg-ag-bf)i$. This is non-trivial because in this setting $a,b$ are encrypted, and we can only use complex multiplication to compute the cross-term.
}
can be represented using complex packing. Furthermore, we can compose any number of Add, Subtract, and Broadcast-Multiply, operations represented using complex packing. Note, real packing supports these operations, as well as Multiply. However, real packing requires
twice the number of slots, i.e. two plaintexts, or doubling $N$.

These properties easily generalize to larger $N$. In essence, complex packing
can perform the same computation (as long as it does not include a
ciphertext-ciphertext Multiply operation) as the real packing representation on
twice as many slots.

Now, following nGraph-HE, \ShortTitle\ uses batch-axis plaintext packing
(Section~\ref{sec:background_difficulties}) during inference to store a 4D
inference tensor of shape $(S, C, H, W)$ (batch size, channels, height, width) a
3D tensor of shape $(C, H, W)$, with each ciphertext packing $S$ values. Each
model weight is stored as a plaintext with the same value in each slot
(encoded using scalar encoding, see Section~\ref{sec:scalarencoding}). Hence, in a neural
network, the FC and Convolution layers consist of only Add, Subtract, and
Broadcast-Multiply operations, suitable for complex packing.
Polynomial activations such as $f(x)=x^2$, in contrast, are not suited for
complex packing since they require ciphertext-ciphertext multiplications.
However, ciphertext-ciphertext multiplications are absent in many neural
networks with ReLU activations. For these networks, complex packing doubles the
throughput.

Kim and Song~\cite{KS18} also propose complex packing, by modifying the
underlying HE scheme. Bergamaschi et al.~\cite{bergamaschi2019homomorphic} use
a similar complex packing idea to train logistic models in a genome-wide
association study (GWAS), with limited speedup due to the requirement of
additional conjugation operations. Our use of complex packing, on the other
hand, applies to neural network inference, and nearly doubles the throughput.

\subsection{CKKS Arithmetic Optimizations}
\label{sec:ckks_arithmetic_optimizations}
We introduce optimizations to ciphertext-plaintext addition and
 multiplication in CKKS, which apply in the
special case of batch-axis packing. A
further ciphertext-plaintext multiplication optimization applies when the
coefficient modulus is less than 32 bits.

\subsubsection{Ciphertext-plaintext Addition}
\label{sec:plain_addition}
Ciphertext-plaintext addition in RNS form requires element-wise addition of two
polynomials in which each sum is reduced with respect to the coefficient modulus
$p_\ell$. With our scalar encoding approach, we instead perform summation of the
same scalar with each element of a polynomial.
Algorithm~\ref{alg:add_cipher_plain_general} shows the ciphertext-plaintext
vector algorithm, compared to Algorithm~\ref{alg:add_cipher_plain_scalar}, which
shows the optimized ciphertext-plaintext scalar addition algorithm. Both
implementations require $O(LN)$ memory and runtime, however
Algorithm~\ref{alg:add_cipher_plain_scalar} is more cache-friendly.

\begin{algorithm}
	\caption{Ciphertext-Plaintext Vector Addition}
	\label{alg:add_cipher_plain_general}
	\begin{algorithmic}[1]
		\Function{Add Cipher-Plain Vector}{$\ct \in \mathcal{C}, \pt \in \mathcal{P}$}
		\For{ $\ell = 1$ to $L$}
			\For{ $n = 1$ to $N$}
				\State $\ct[\ell][n] \gets (\ct[\ell][n] + \pt[\ell][n]) \text{\ mod\ } p_\ell$
			\EndFor
		\EndFor
		\EndFunction
	\end{algorithmic}
\end{algorithm}

\begin{algorithm}
	\caption{Ciphertext-Plaintext Scalar Addition}
	\label{alg:add_cipher_plain_scalar}
	\begin{algorithmic}[1]
		\Function{Add Cipher-Plain Scalar}{$\ct \in \mathcal{C}, \pt \in \mathcal{P}$}
		\For{ $\ell = 1$ to $L$}
		\State $tmp \gets \pt[\ell]$
			\For{ $n = 1$ to $N$}
				\State $\ct[\ell][n] \gets (\ct[\ell][n] +tmp) \text{\ mod\ } p_\ell$
			\EndFor
		\EndFor
		\EndFunction
\end{algorithmic}
\end{algorithm}

Note that the same optimization works for ciphertext-plaintext subtraction, and we expect
similar improvements.

\subsubsection{Ciphertext-plaintext Multiplication}
\label{sec:plain_mult}
Ciphertext-plaintext multiplication in RNS form requires element-wise multiplication of
two polynomials in which each product is reduced with respect to the coefficient
modulus $q_l$. The modulus reduction is performed with Barrett
reduction~\cite{barrett1986implementing}. We present two optimizations.

First, our scalar encoding allows us to perform multiplication between a scalar
and each element of the polynomial, rather than between two polynomials. This is
the same optimization as in ciphertext-plaintext addition.

Second, we provide an optimization for the case in which the coefficient modulus
is 32 bits, rather than 64 bits. The benefit arises from a simpler
implementation of Barrett reduction which requires fewer additions and
multiplications. In SEAL, ciphertext and plaintext elements are stored at 64-bit
unsigned integers, with a maximum modulus of 62 bits~\cite{sealcrypto}. As a
result, performing the multiplication may overflow to 128 bits. Then, performing
Barrett reduction requires 5 multiplications, 6 additions, and 2 subtractions (including the
conditional subtraction). See Algorithm~\ref{alg:barrett64} for the pseudocode,
which closely follows SEAL's implementation\footnote{\url{https://github.com/microsoft/SEAL/blob/3.3.0/native/src/seal/util/uintarithsmallmod.h\#L146-L187}}.
We store an unsigned 128-bit number $z$ as two unsigned 64-bit
numbers with $z[0]$ containing the 64 low bits and $z[1]$ containing the 64 high
bits. The $add64$ function will return the carry bits of the addition.
\begin{algorithm}
	\algnewcommand{\LineComment}[1]{\State \(\triangleright\) #1}
	\caption{BarrettReduction128}
	\label{alg:barrett64}
	\begin{algorithmic}[1]
		\Function{BarrettReduction128}{128-bit number $z$, 64-bit modulus $q$, 128-bit Barrett ratio $r$}
	\State $uint64\ tmp1, tmp2[2],	tmp3, carry$
	\State $carry \gets mult\_hw64(z[0], r[0])$ \Comment{Multiply low bits}
	\State $tmp2 \gets z[0] * r[1]$
    \LineComment{Compute high bits of $z[0] * r$}
	\State $tmp3 \gets tmp2[1] + add64(tmp2[0], carry,\&tmp1)$
	\State $tmp2 \gets z[1] * r[0]$
	\State $carry \gets tmp2[1] + add64(tmp1,tmp2[0], \&tmp1)$
	\State $tmp1 \gets z[1] * r[1] + tmp3 + carry$	\Comment{Compute $[z * r]_{2^{128}}$}
	\State $tmp3 \gets z[0] - tmp1 * q$ 	\Comment{Barrett subtraction}
	\If{$tmp3 \geq q$}  \Comment{Conditional Barrett subtraction}
		\State $result \gets tmp3 - q$
	\Else
		\State $result	\gets tmp3$
	\EndIf
	\State \Return $result$
	\EndFunction
	\end{algorithmic}
\end{algorithm}

In the case where $q$ is a 32-bit modulus, the Barrett reduction becomes much
simpler, requiring just 2 multiplications and 2 subtractions (including the
conditional subtraction). Algorithm~\ref{alg:barrett32} shows the pseudocode for
the more efficient Barrett reduction, which closely follows SEAL's implementation\footnote{\url{https://github.com/microsoft/SEAL/blob/3.3.0/native/src/seal/util/uintarithsmallmod.h\#L189-L217}}. Algorithm~\ref{alg:mult_cipher_plain_general} shows the general, 64-bit modulus implementation of ciphertext-plaintext multiplication. Note, SEAL uses Barrett64 reduction for rescaling, whereas we use it for optimized ciphertext-plaintext multiplication.

Algorithm~\ref{alg:mult_cipher_plain_scalar32} shows the optimized 32-bit
modulus implementation of multiplication with a scalar plaintext. Note, the
plaintext $\pt$ contains only $L$ entries, rather than $N \cdot L$ entries. Algorithm~\ref{alg:mult_cipher_plain_general} and Algorithm~\ref{alg:mult_cipher_plain_scalar32} both require $O(LN)$ runtime; however, Algorithm~\ref{alg:mult_cipher_plain_scalar32} is more cache-friendly.

\begin{algorithm}
	\caption{BarrettReduction64}
	\label{alg:barrett32}
	\begin{algorithmic}[1]
		\Function{BarrettReduction64}{64-bit number $z$, 32-bit
			modulus $q$, 64-bit Barrett ratio $r$}
		\State $uint64\ carry$
		\State $carry \gets mult\_hw64(z, r)$
		\Comment{Compute $[z \cdot q ]_{2^{64}}$}
		\State $carry \gets z - carry * q$ \Comment{Barrett subtraction}
		\If{$carry \geq q$}  \Comment{Conditional Barrett subtraction}
		\State $result \gets carry -	q$
		\Else
		\State $result \gets carry$
		\EndIf
		\State \Return $result$
		\EndFunction
	\end{algorithmic}
\end{algorithm}
\begin{algorithm}
	\caption{Ciphertext-Plaintext 64-bit Multiplication}
	\label{alg:mult_cipher_plain_general}
	\begin{algorithmic}[1]
		\Function{Multiply Cipher-Plain 64-bit}{$\ct \in \mathcal{C}, \pt \in \mathbb{Z}^{L \times N}$, 128-bit Barrett ratio $r$}
		\For{$\ell = 1$ to $L$}
		\For{ $n = 1$ to $N$}
		\State $uint64\  z[2];$
		\State $z \gets \ct[\ell][n] * \pt[\ell][n]$ \Comment{Perform multiplication}
		\State $\ct[\ell][n] \gets BarrettReduction128(z, q_\ell, r)$
		\EndFor
		\EndFor
		\EndFunction
	\end{algorithmic}
\end{algorithm}
\begin{algorithm}
	\caption{Ciphertext-Plaintext Scalar 32-bit Multiplication}
	\label{alg:mult_cipher_plain_scalar32}
	\begin{algorithmic}[1]
		\Function{Multiply Cipher-Plain 32-bit}{$\ct \in \mathcal{C}, \pt \in \mathbb{Z}^L$, 64bit Barrett ratio $r$}
		\For{ $\ell = 1$ to	$L$}
		\State $tmp \gets \pt[\ell]$
			\For{ $n = 1$ to $N$}
				\State $uint64\  z;$
				\State $z \gets \ct[\ell][n]* tmp$ \Comment{Perform multiplication}
				\State $\ct[\ell][n] \gets BarrettReduction64(z, q_\ell, r)$
			\EndFor
		\EndFor
		\EndFunction
	\end{algorithmic}
\end{algorithm}
\subsection{Graph-level Optimizations}
\label{sec:graph_opt}
In addition to the above low-level CKKS optimizations, we present two graph-level optimizations.

\subsubsection{Lazy Rescaling}
Rescaling in CKKS can be thought of as a procedure which homomorphically removes the
inaccurate LSBs in the (encrypted) message. See
Section~\ref{sec:math_background} for a longer description,
or~\cite{cheon2017homomorphic} for full details. Due to the NTT and NTT$^{-1}$,
rescaling is ${\approx}9$x more expensive than ciphertext-plaintext multiplication
in SEAL (see Appendix~\ref{app:seal_perf_test}). The \emph{naive rescaling}
approach rescales after every multiplication. \emph{Lazy rescaling},
on the other hand, minimizes the number of rescaling operations by:
\begin{itemize}
	\item rescaling only after a Fully-Connected (FC) or Convolution layer, rather than
	after every multiplication therein;
	\item skipping rescaling if there are no subsequent multiplications before the
	ciphertext is decrypted.
\end{itemize}
Since FC and Convolution layers each contain several multiplications per output
element, the first optimization reduces the number of rescaling operations performed by a factor of the inner dimension (for FC layers) or window size (for Convolution
layers).

The second optimization ensures rescaling happens only when
reducing the message scale is necessary. In particular, addition is allowed
post-multiplication, pre-rescaling. In the case where the last two layers of a
network (or before a ReLU activation in a hybrid MPC-HE scheme, see
Section~\ref{sec:fixed_network}) are FC-Add, Convolution-Add or Multiply-Add,
this ensures the rescaling is omitted entirely. Note, for a choice of parameters
with $L = L_f$, where $L_f$ is the multiplicative depth of the function, this
optimization is equivalent to skipping rescaling to level 0.
Table~\ref{tab:rescaling_example} shows an example where the second optimization
results in a skipped rescaling. Note, lazy rescaling applies `Use rescaling
sparingly' from~\cite{blatt2019optimized}, to neural network inference instead
of a genome-wide association study (GWAS). The GWAS setting has a closed-form
semi-parallel logistic regression model, whereas our setting involves long sequences of linear and non-linear operations on tensors, e.g. convolutions, and pooling operations.

\begin{table}
	\begin{center}
		\caption{Benefit of lazy rescaling at level 0. Lazy rescaling skips the rescaling, whereas naive rescaling performs an unnecessary rescaling.}
		\label{tab:rescaling_example}
		\begin{tabular}{l c c c c  c c}
			\toprule
			\multirow{2}{*}{\textbf{Operation}} &
			\multicolumn{2}{c}{\textbf{Number of rescalings}} \\
			\cmidrule{2-3}
			& \textbf{Naive Rescaling} & \textbf{Lazy Rescaling}  \\ \midrule
			Constant & $L-1$ & $L-1$ \\
			Multiply & $L$ & $L-1$ \\
			Add & $L$ & $L-1$ \\
			\bottomrule
		\end{tabular}
	\end{center}
\end{table}

\subsubsection{Depth-aware Encoding}
The runtime complexity and memory usage of encoding a scalar at level $\ell$ in
SEAL are both $O(N \ell)$ (see Section~\ref{sec:scalarencoding}). Throughout the
execution of HE operations, the level $\ell$ decreases due to the rescaling
operation (see Section~\ref{sec:rescaling}). When multiplying or adding a
ciphertext $\ct$ at level $\ell < L$ with a plaintext $\pt$, it is therefore
advantageous to encode $\pt$ at level $\ell$ rather than level $L$, as noted by the `Harnessing the CRT ladder' technique in~\cite{blatt2019optimized}. This will
reduce both the runtime and memory usage of the encoding step. In practice, this
implementation can have two forms:
\begin{enumerate}
	\item
		\emph{Compile-time encoding.} An optimization pass through the computation
		graph can identify the level at which each plaintext is encoded. This
		compilation step requires a larger initial memory overhead, for the benefit
		of increased runtime.
	\item
	 \emph{Lazy encoding}. In this implementation, the plaintext model weights are
	 stored in native scalar (i.e. floating-point) format, and encoding is delayed
	 until immediately preceding multiplication or addition with a ciphertext $\ct$.
	 The level $\ell$ at which to encode the model weight is determined by
	 observing the level of $\ct$.
\end{enumerate}
If encoding is expensive compared to addition/multiplication (as in SEAL, see
Appendix~\ref{app:seal_perf_test}), compile-time encoding yields the fastest
runtime. However, due to the choice batch-axis packing, \ShortTitle's
scalar encoding (Section~\ref{sec:scalarencoding}) is significantly cheaper than
addition/multiplication, requiring runtime and memory $O(\ell)$, compared to
$O(\ell N \log N)$ runtime and $O(\ell N)$ memory usage of general encoding.
Hence, performing lazy encoding at runtime results in little slowdown, and
allows for a simpler implementation.

Here, we introduced CKKS-specific optimizations to scalar encoding, ciphertext-plaintext addition, and ciphertext-plaintext multiplication in the batch-axis packing case. We also introduced graph-level optimizations of complex packing and depth-aware encoding.

%% file: results.tex
\section{Evaluation}
We evaluate our optimizations on small, single-operation tests
(Section~\ref{sec:result_low_level_operations}), as well as on larger neural networks
(Section~\ref{sec:nn_evaluation}). All results are computed on Intel
Xeon\textregistered\ Platinum 8180 2.5GHz systems with 376GB of RAM and 112 cores, running Ubuntu 16.04. The
localhost bandwidth is
\SI[per-mode=symbol]{46.2}{\giga\bit\per\second}, and the local area network
(LAN) bandwidth is \SI[per-mode=symbol]{9.4}{\giga\bit\per\second}. We use GCC
7.4 with -O2 optimization.




\subsection{Client-aided Model}
\label{sec:client_server}
To mitigate the classification accuracy degradation of HE-friendly networks
(Section~\ref{sec:he_for_deep_learning}), we implement a simple two-party computation
approach. Specifically, evaluate a non-polynomial function
$f$ on a ciphertext $\ct$, the server sends $\ct$ to the client, which decrypts
$Dec(\ct) \rightarrow \pt$, computes $f(\pt)$ , and sends a fresh encryption of
the result, $Enc(f(\pt))$ to the server. This approach accomplishes two tasks:
first, it enables the computation of non-polynomial functions; second, it
refreshes the ciphertext, i.e. resets the noise budget and coefficient modulus
to the highest level $L$. However, this approach can leak information about the
model to the client, as it provides the pre-activation values $\pt$ to the client, as well as the activation function itself.
One possible improvement is performing the non-polynomial function using additive masking and
garbled circuits, as in Gazelle~\cite{juvekar2018gazelle}. Another approach is
to perform the decryption, non-polynomial activation, and encryption in a
trusted execution environment (TEE) attested to the user, such as Intel's Software Guard Extensions
(SGX)~\cite{graphene}. For instance, Wang et. al~\cite{wang2019toward} use Intel's SGX for bootstrapping
only, though this approach is easily adapted to perform the non-polynomial
activation as well.

Our client-aided approach, therefore, represents a placeholder for more secure
implementations in future work. 
Since the client-aided model refreshes ciphertexts at each non-polynomial
layer, the effective multiplicative depth is reduced to the multiplicative depth
between non-polynomial layers. This enables the computation of arbitrarily-deep
neural networks with much smaller encryption parameters, and therefore much
faster runtimes.

\subsection{Low-level Operations}
\label{sec:result_low_level_operations}
\subsubsection{Scalar Encoding}
\label{sec:result_scalar_encoding}
We implement the scalar encoding optimization from
Section~\ref{sec:scalarencoding}. Table~\ref{tab:encoding} shows the speedup of
several parameter choices, each satisfying $\lambda=128$-bit security. As
expected, the memory improvement is a factor of $N$. The runtime improvement of
scalar encoding increases with $N$, due to the $O(L)$ runtime, compared to
$O(LN \log N)$ in the general encoding.

\begin{table}
	\begin{center}
		\small
		\renewcommand\theadfont{\small}
		\caption{Runtime and memory usage when encoding a scalar, with and without optimization (Opt.). Runtimes are averaged across 1000 trials.}
		\label{tab:encoding}
		\begin{tabular}{c c c
				S[separate-uncertainty,
				table-number-alignment = center,
				table-figures-integer = 7,
				table-figures-decimal = 0,
				table-figures-uncertainty = 0]
				S[separate-uncertainty,
				table-number-alignment = center,
				table-figures-integer = 4,
				table-figures-decimal = 0,
				table-figures-uncertainty = 0]
				S[separate-uncertainty,
				table-number-alignment = center,
				table-figures-integer = 5,
				table-figures-decimal = 0,
				table-figures-uncertainty = 0]
				S[separate-uncertainty,
				table-number-alignment = center,
				table-figures-integer = 2,
				table-figures-decimal = 1,
				table-figures-uncertainty = 0]
			}
			\toprule
			\multirow{2}{*}{$\boldsymbol N$} &
			\multirow{2}{*}{$\boldsymbol L$} &
			\multirow{2}{*}{\textbf{Opt.}} &
			\multicolumn{2}{c}{\textbf{Memory}} &
			\multicolumn{2}{c}{\textbf{Runtime}} \\
			\cmidrule(lr){4-5}\cmidrule(l){6-7}
			& & & \textbf{Usage (bytes)} & \textbf{Improv.} & \textbf{Time ($\ns$)} &
			\textbf{Speedup} \\
			 \midrule

		$2^{12}$ & 1 & \xmark & 32768 & & 605   \tabularnewline
		$2^{12}$ & 1 &  \cmark & 8 & 4096 & 177 & 3.4  \tabularnewline \midrule
		$2^{13}$ & 3 & \xmark &  196608 & & 2951  \tabularnewline
		$2^{13}$ & 3 & 	\cmark & 24 & 8192 & 202 & 14.6  \tabularnewline 	\midrule
		$2^{14}$ & 8 & \xmark & 1048576 & & 38938 \tabularnewline
		$2^{14}$ & 8 & \cmark & 64 & 16384 & 443 & 87.9	\tabularnewline
			\bottomrule
		\end{tabular}
	\end{center}
\end{table}

\subsubsection{Ciphertext-plaintext Addition}
We implement the scalar addition optimization from Section~\ref{sec:plain_addition}.
Table~\ref{tab:plain_addition} shows a speedup of \tocheck{2.6x-4.2x}, with more
speedup for larger encryption parameters. Algorithm~\ref{alg:add_cipher_plain_general} and
Algorithm~\ref{alg:add_cipher_plain_scalar} both have $O(LN)$ runtime
complexity. The primary source of speedup, therefore, is due to the fact that one of the
two operands, the scalar, is kept into the processor registers, and the other operand
does not have to compete for its placement and retrieval from the cache memory.

\begin{table}[!ht]
	\begin{center}
		\caption{Runtime improvement in ciphertext-plaintext scalar addition. Parameter choices satisfy $\lambda=128$-bit security. Runtimes are averaged across 1000 trials.}
		\label{tab:plain_addition}
\begin{tabular}{ c c
    S[separate-uncertainty,
		table-number-alignment = center,
		table-figures-integer = 3,
		table-figures-decimal = 1,
		table-figures-uncertainty = 0]
	S[separate-uncertainty,
		table-number-alignment = center,
		table-figures-integer = 2,
		table-figures-decimal = 1,
		table-figures-uncertainty = 0]
		 c c }
	\toprule
	\multirow{2}{*}{$\boldsymbol N$} &
	\multirow{2}{*}{$\boldsymbol L$} &
	\multicolumn{2}{c}{\textbf{Runtime ($\mics$)}} \\
	\cmidrule(lr){3-4}
	& & \textbf{General}
	& \textbf{Scalar} &
	\centering{\textbf{Speedup}} \tabularnewline
	\midrule
	$2^{12}$ & 1 & 2.3 & 0.9 & 2.6 \tabularnewline
	$2^{13}$ & 3 & 12.6 &4.5 & 2.8 \tabularnewline
	$2^{14}$ & 8 & 124.5 & 30.0 & 4.2 \tabularnewline
	\bottomrule
		\end{tabular}
	\end{center}
\end{table}

\subsubsection{Ciphertext-plaintext Multiplication}
We implement the scalar multiplication optimization from
Section~\ref{sec:plain_mult}. Table~\ref{tab:plain_mult} shows a speedup of
\tocheck{2.6}x for different parameter choices. Notably, the parameters uses 30-bit
coefficient moduli, so our Barrett reduction optimization applies.

\begin{table}[!ht]
	\renewcommand\theadfont{\normalsize}
	\begin{center}
		\caption{Runtime improvement in ciphertext-plaintext scalar multiplication. Parameter choices satisfy $\lambda=128$-bit security. Runtimes are averaged across 1000 trials.}
		\label{tab:plain_mult}
\begin{tabular}{ c c c
	   S[separate-uncertainty,
		table-number-alignment = center,
		table-figures-integer = 3,
		table-figures-decimal = 1,
		table-figures-uncertainty = 0]
		 c c }
	\toprule
	\multirow{2}{*}{$\boldsymbol N$} & \multirow{2}{*}{$\boldsymbol L$} & \multicolumn{2}{c}{\textbf{Runtime ($\mics$)}} \\
	\cmidrule(lr){3-4}
	& & \centering \textbf{General} &
	\textbf{Scalar} &
	\centering{\textbf{Speedup}}
	\tabularnewline	\midrule
	$2^{13}$ & 3 & 181.7 & 71.1 & 2.6 \tabularnewline
	$2^{14}$ & 8 & 966.6 & 377.6 & 2.6 \tabularnewline
	\bottomrule
		\end{tabular}
	\end{center}
\end{table}

\subsection{Neural Network Workloads}
\label{sec:nn_evaluation}
To evaluate our graph-level optimizations and complex
packing, we evaluate two neural networks: the standard
CryptoNets~\cite{gilad2016cryptonets} model, and MobileNetV2~\cite{sandler2018mobilenetv2}. To the best of our
knowledge, this is the largest network whose linear layers have been
homomorphically evaluated, as well as the first homomorphic
evaluation of a network on the ImageNet dataset.

\subsubsection{CryptoNets}
\label{sec:cryptonets_evaluation}
The CryptoNets network~\cite{gilad2016cryptonets} is the seminal HE-friendly DL
model for the MNIST handwritten digits dataset~\cite{lecun1998mnist}. The
architecture uses $f(x) = x^2$ for the activations, and has a multiplicative
depth of 5. See Appendix~\ref{app:architectures} for the full architecture. As
in~\cite{gilad2016cryptonets}, we achieve 98.95\percent\
accuracy. Table~\ref{tab:cryptonets} shows lazy rescaling reduces the
runtime of the CryptoNets network by
\tocheck{}${\approx}8$x. Multi-threading further improves the performance (see Appendix~\ref{app:parallel_scaling}).
\begin{table}
	\begin{center}
		\renewcommand\theadfont{\normalsize}
		\caption{Impact of lazy rescaling on CryptoNets runtime using $N=2^{13}, L=6$, with accuracy 98.95\percent.}
		\label{tab:cryptonets}
		\begin{tabular}{
				S[separate-uncertainty,
				table-number-alignment = center,
				table-figures-integer = 2,
				table-figures-decimal = 0,
				table-figures-uncertainty = 0]
				 c
				S[separate-uncertainty,
				table-number-alignment = center,
				table-figures-integer = 2,
				table-figures-decimal = 2,
				table-figures-uncertainty = 0]
				S[separate-uncertainty,
				table-number-alignment = center,
				table-figures-integer = 4,
				table-figures-decimal = 2,
				table-figures-uncertainty = 3]}
			\toprule
			{\centering \multirow{2}{*}{ \thead{\textbf{Thread} \\ \textbf{Count}}}} &
			{\centering \multirow{2}{*}{ \thead{\textbf{Lazy} \\ \textbf{Rescaling}}}} &
			\multicolumn{2}{c}{\textbf{Runtime}} \\ \cmidrule(l){3-4}
			&& \textbf{Amortized (\ms)} &
			{\centering \textbf{Total (\Sec)}} \\ \midrule
			1 & \xmark & 59.21 & 242.51 \pm 3.69 \tabularnewline
			1 & \cmark & 7.23 & 29.62 \pm 0.63 \tabularnewline
			24 & \cmark & 0.50 & 2.05 \pm 0.11 \tabularnewline
			\bottomrule
		\end{tabular}
	\end{center}
\end{table}

In order to show the benefits of complex packing
(Section~\ref{sec:complex_packing}), we implement the client-aided model
(Section~\ref{sec:client_server}). We train the CryptoNets network with
ReLU activations rather than $x^2$ activations, and add bias terms.
See Appendix~\ref{app:architectures} for the complete architecture. The use of ReLU
activations effectively decreases the multiplicative depth to 1, since the
client-aided computation of ReLU refreshes the ciphertexts. This lower multiplicative depth enables much smaller encryption parameters, $N=2^{11}, L=1$, with a single 54-bit coefficient modulus.

Table~\ref{tab:cryptonets_relu} shows the improvement due to complex packing.
Complex packing does not take advantage of our scalar encoding optimization
(Section~\ref{sec:scalarencoding}), slightly increasing the runtime from the real
packing case. Nevertheless, complex packing roughly halves the amortized runtime
by doubling the capacity.

The total runtime is much smaller than the runtime in
Table~\ref{tab:cryptonets}, due to the use of much smaller encryption
parameters. The amortized runtime is also improved, though less dramatically. Note, the communication overhead between the server and client accounts for roughly
\tocheck{27}\percent\ of the runtime in the LAN setting. Optimizing the communication leaves room for future improvement.

\begin{table}
	\begin{center}
		\footnotesize
		\renewcommand\theadfont{\footnotesize}
		\caption{Impact of complex packing on CryptoNets with ReLU activations using $N=2^{11}, L=1$, and 98.62\percent\  accuracy. Results are averaged over 10 trials. Amt. times are amortized over the largest batch size supported.}
		\label{tab:cryptonets_relu}
		\begin{tabular}{S[separate-uncertainty,
				table-number-alignment = center,
				table-figures-integer = 2,
				table-figures-decimal = 0,
				table-figures-uncertainty = 0]
			 c
			 c
			 S[separate-uncertainty,
			 table-number-alignment = center,
			 table-figures-integer = 4,
			 table-figures-decimal = 0,
			 table-figures-uncertainty = 0]
			 c
			S[separate-uncertainty,
			table-number-alignment = center,
			table-figures-integer = 1,
			table-figures-decimal = 2,
			table-figures-uncertainty = 3]
		}
			\toprule
			{\centering \multirow{2}{*}{ \thead{\textbf{Thread} \\ \textbf{Count}}}} &
			{\centering \multirow{2}{*}{\small \thead{\textbf{Complex} \\ \textbf{packing}}}} &
			{\centering \multirow{2}{*}{\small \thead{\textbf{Network} \\ \textbf{setting}}}} &
			{\centering \multirow{2}{*}{ \thead{\textbf{Batch} \\ \textbf{size}}}} &
			\multicolumn{2}{c}{\textbf{Runtime}} \\ \cmidrule(l){5-6}
			& & & & \textbf{Amt. ($\ms$)} &
			{\centering \textbf{Total ($\Sec$)}} \\ \midrule
			1 & \xmark & localhost & 1024 & 2.72 & 2.79 \pm 0.06 \tabularnewline
			1 & \cmark & localhost & 2048 & 1.44 & 2.94 \pm 0.04 \tabularnewline
			24 & \cmark & localhost & 2048 & 0.24 & 0.5 \pm 0.04 \tabularnewline
			24 & \cmark & LAN & 2048 & 0.34 & 0.69 \pm 0.04 \tabularnewline
			\bottomrule
		\end{tabular}
	\end{center}
\end{table}

Table~\ref{tab:cryptonets_perf_cmp} shows the performance of \ShortTitle\ on the
CryptoNets network compared to existing methods. Lola and Gazelle optimize for
latency at the cost of reduced throughput. Other methods, such as CryptoNets,
Faster CryptoNets, and nGraph-HE adopt the same batch-axis packing as we do,
thereby optimizing for throughput. Our method achieves the highest throughput of
$\SI{1998}{\imagespersec}$ on the CryptoNets network. Furthermore, the
client-aided model enables an even higher throughput of
$\SI{2959}{\imagespersec}$. Notably, the latency of our approach is much smaller
than previous batch-axis packing approaches, and has similar runtime as the
latency-optimized LoLa, while achieving much larger throughput.

\begin{table}
	\begin{center}
		\newcommand{\tbnum}[1]{\multicolumn{1}{c}{\bfseries \num{#1}}}

		\footnotesize
		\renewcommand\theadfont{\footnotesize}
		\caption{CryptoNets performance comparison, including accuracy (Acc.), latency (Lat.), and throughput (Thput.). For hybrid protocols, latency is reported in the LAN setting and communication (Comm.) includes only the interactive part of the protocol.}
		\label{tab:cryptonets_perf_cmp}
		\begin{tabular}{l
			   S[separate-uncertainty,
				table-number-alignment = center,
				table-figures-integer = 2,
				table-figures-decimal = 2,
				table-figures-uncertainty = 0]
				S[separate-uncertainty,
				table-number-alignment = center,
				table-figures-integer = 3,
				table-figures-decimal = 2,
				table-figures-uncertainty = 0]
				S[separate-uncertainty,
				table-number-alignment = center,
				table-figures-integer = 4,
				table-figures-decimal = 1,
				table-figures-uncertainty = 0]
				c
				S[separate-uncertainty,
				table-number-alignment = center,
				table-figures-integer = 2,
				table-figures-decimal = 2,
				table-figures-uncertainty = 0]
				c
				c
			}
			\toprule
			\multirow{2}{*}{\textbf{Method}} &
			{\centering \multirow{2}{*}{ \thead{\textbf{Acc.} \\ (\percent)}}} &
			{\centering \multirow{2}{*}{ \thead{\textbf{Lat.} \\ ($\Sec$)}}} &
			{\centering \multirow{2}{*}{ \thead{\textbf{Thput.} \\ (\impersec)}}} &
			\multirow{2}{*}{\textbf{Protocol}} &
			{\centering \multirow{2}{*}{ \thead{\textbf{Comm.} \\ (\mbperim)}}} \\ \\
			\midrule
			LoLa~\cite{Brutzkus2019LowLatency} & 98.95 & 2.2 & 0.5 & HE &  \tabularnewline

			FHE-DiNN100~\cite{bourse2018fast} & 96.35 & 1.65 & 0.6 & HE & 	\tabularnewline

			CryptoNets~\cite{gilad2016cryptonets} & 98.95 & 250 & 16.4 & HE &  \tabularnewline

			Faster CryptoNets~\cite{chou2018faster} & 98.7 & 39.1 & 210 & HE &
			\tabularnewline

			nGraph-HE~\cite{boemer2019ngraph} & 98.95 & 16.7 & 245 & HE &  \tabularnewline


			CryptoNets 3.2~\cite{Brutzkus2019LowLatency} & 98.95 & 25.6 & 320 & HE &  \tabularnewline


			\ShortTitle & 98.95 & 2.05 & 1998 & HE &  \tabularnewline

			 \midrule


			Chameleon~\cite{riazi2018chameleon} & 99 & 2.24 &  1.0 & HE-MPC & 5.1  \tabularnewline

			MiniONN~\cite{minionn} & 98.95 & 1.28 & 2.4 & HE-MPC & 44 \tabularnewline

			Gazelle~\cite{juvekar2018gazelle} & 98.95 & 0.03 & 33.3 & HE-MPC & 0.5 \tabularnewline

			\ShortTitle-ReLU & 98.62 &  0.69 &  2959 & HE-MPC &  0.03  \tabularnewline

			\bottomrule
		\end{tabular}
	\end{center}
\end{table}

\subsubsection{MobileNetV2}
\label{sec:mobiletnetv2}
ImageNet~\cite{deng2009imagenet} is a dataset used for image recognition, consisting of colored images, ${\approx}1.2$ million for training, and \SI{50000} for
validation, classified into \SI{1000} categories.
The images vary in size, though they are commonly rescaled to shape $224
\times 224 \times 3$. The large number of categories and
large image resolution make ImageNet a much more difficult task than MNIST or
CIFAR10~\cite{krizhevsky2014cifar}. MobileNetV2~\cite{sandler2018mobilenetv2} is a
lightweight network architecture which achieves high accuracy on ImageNet with a
small number of parameters and multiply-accumulate operations. MobileNets are
parameterized by an expansion factor, which can be used to reduce the model
size, resulting in a faster runtime at the expense of lower accuracy. The ReLU activations reduce the effective multiplicative depth, enabling use of
small encryption parameters, $N=2^{12}$ and $L=3$ coefficient moduli at
$\lambda=128$-bit security. Furthermore, the lack of ciphertext-ciphertext
multiplications enables use of complex packing. We demonstrate \ShortTitle\ on
MobileNetV2 with expansion factor 0.35, and image ranging from size
$96 \times 96$ to the full size, $224 \times 224$.

\begin{figure}
	\centering
	\includegraphics[width=\columnwidth]{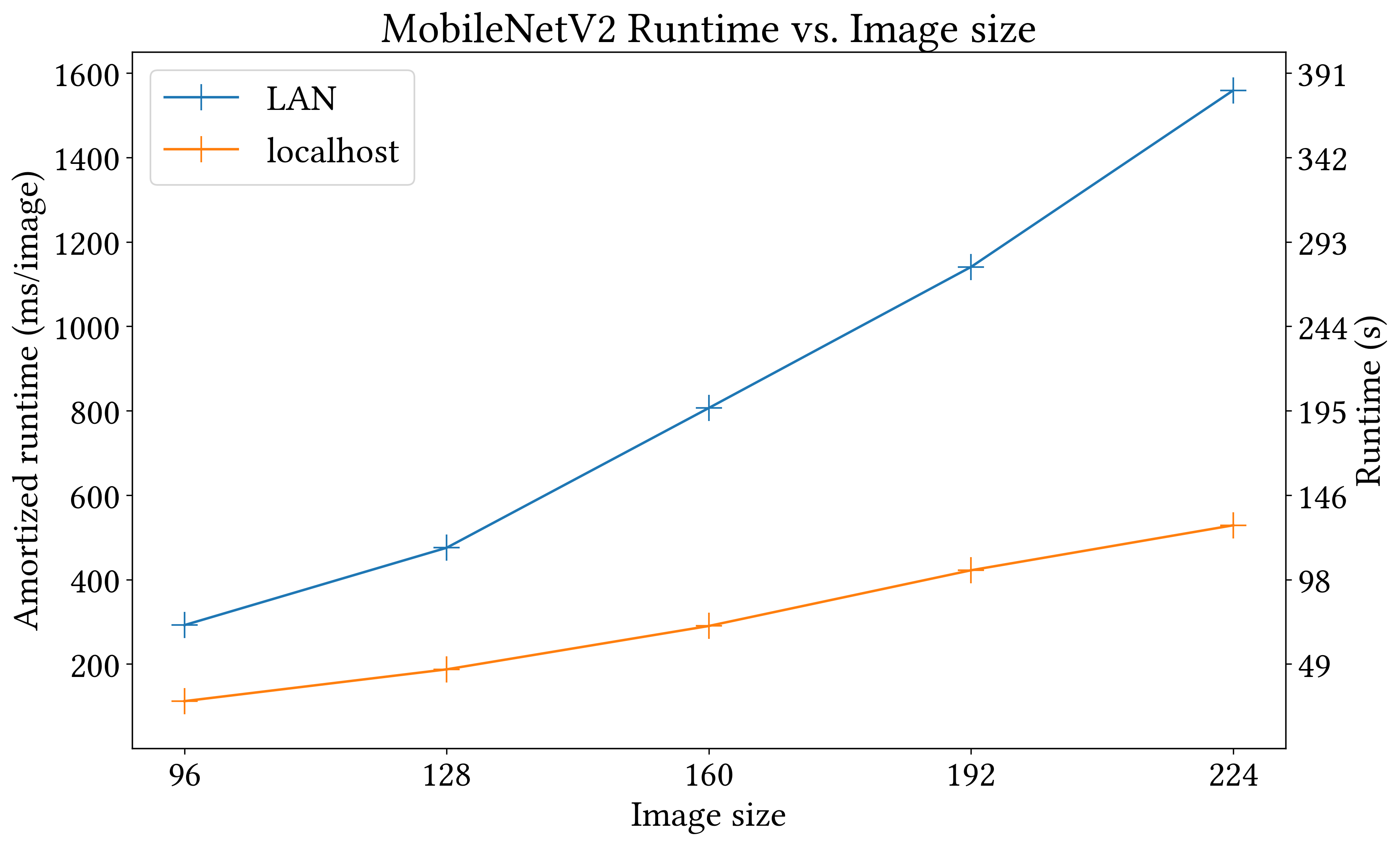}
	\caption{Runtime vs. Image size of LAN and localhost MobileNetV2 models. Table~\ref{tab:mobilenet} shows the corresponding accuracies.}
	\label{fig:mobilenet_wan_vs_lan}
\end{figure}

Table~\ref{tab:mobilenet} shows the results from MobileNetV2 inference on a
variety of image sizes. The large increase in runtime from the localhost setting
to the LAN setting is due to the communication overhead. The localhost setting
therefore represents a lower-bound to the timings possible in the LAN setting.
Notably, the accuracy degradation due to HE is ${\approx}$0.01\percent, less than 7 images in \SI{50000}{}. Figure~\ref{fig:mobilenet_wan_vs_lan} shows the increase in runtime with larger images sizes, and the significant latency introduced by the LAN setting.

\ifiacr
\else 
\begin{table*}
	\caption{MobileNetV2 results on localhost and LAN settings using complex packing, batch size 4096, 56 threads, and encryption parameters $N=2^{12}, L=3$ at $\lambda=128$-bit security. Runtimes are averaged across 10 trials. Encrypting the data reduces the top-1 accuracy by an average of  0.0136\percent, ${\approx} 7$ images in \SI{50000}{}.}
	\label{tab:mobilenet}
	\centering
	\footnotesize
	\renewcommand\theadfont{\footnotesize}
	\begin{tabular}{c c c
			r
			r
			S[separate-uncertainty,
			table-number-alignment = center,
			table-figures-integer = 3,
			table-figures-decimal = 0,
			table-figures-uncertainty = 0]
			 S[separate-uncertainty,
			 table-number-alignment = center,
			 table-figures-integer = 3,
			 table-figures-decimal = 0,
			 table-figures-uncertainty = 2]
			 S[separate-uncertainty,
			 table-number-alignment = center,
			 table-figures-integer = 3,
			 table-figures-decimal = 0,
			 table-figures-uncertainty = 0]
			 S[separate-uncertainty,
			 table-number-alignment = center,
			 table-figures-integer = 4,
			 table-figures-decimal = 0,
			 table-figures-uncertainty = 2]
		 	S[separate-uncertainty,
		 	table-number-alignment = center,
		 	table-figures-integer = 3,
		 	table-figures-decimal = 0,
		 	table-figures-uncertainty = 1]
	 		S[separate-uncertainty,
	 		table-number-alignment = center,
	 		table-figures-integer = 2,
	 		table-figures-decimal = 0,
	 		table-figures-uncertainty = 1]
	 		S[separate-uncertainty,
	 		table-number-alignment = center,
	 		table-figures-integer = 3,
	 		table-figures-decimal = 0,
	 		table-figures-uncertainty = 1]}
		\toprule
		{\centering \multirow{4}{*}{\thead{{ }\\ \textbf{MobileNetV2} \\ \textbf{Model} \\ { }}}} &
		\multicolumn{2}{c}{
			\multirow{2}{*}{\thead{\textbf{Unencrypted} \\ \textbf{Accuracy (\percent)}}}} &
		\multicolumn{2}{c}{
			\multirow{2}{*}{\thead{\textbf{Encrypted} \\ \textbf{Accuracy (\percent)}}}} &
		\multicolumn{4}{c}{\textbf{Runtime}} &
		{\centering \multirow{3}{*}{\thead{{ }\\ \textbf{Communication} \\ (\mbperimage)}}} &

		\multicolumn{2}{c}{
			\multirow{2}{*}{\thead{\textbf{Memory} \\ \textbf{(\giga\byte)}}}} \\

		\cmidrule(l){6-9} & & & & &
		\multicolumn{2}{c}{\textbf{Localhost}} &
		\multicolumn{2}{c}{\textbf{LAN}}  \\
		\cmidrule(lr){2-3} 	\cmidrule(lr){4-5}
		\cmidrule(lr){6-7} \cmidrule(l){8-9}  \cmidrule(l){11-12} &
		{\centering \textbf{Top-1}} &
		{\centering \textbf{Top-5}} &
		\multicolumn{1}{c}{\textbf{Top-1}} &
		\multicolumn{1}{c}{\textbf{Top-5}} &
		\textbf{Amt. ($\ms$)} &
		{\centering \textbf{Total ($\Sec$)}} &
		\textbf{Amt. ($\ms$)} &
		{\centering \textbf{Total ($\Sec$)}} &
		&
		{\centering \textbf{Client}} &
		{\centering \textbf{Server}}
		\tabularnewline
		\midrule

0.35-96 & 42.370 & 67.106 & $42.356$ $(-0.014)$ & $67.114$ $(+0.008)$ & 27 & 112 \pm 5 & 71 & 292 \pm 5 & 38.4 & 8.6 & 60.3 \tabularnewline
0.35-128 & 50.032 & 74.382 & $49.982$ $(-0.050)$ & $74.358$ $(-0.024)$ & 46 & 187 \pm 4 & 116 & 475 \pm 10  & 63.7 & 12.6 & 100.4 \tabularnewline
0.35-160 & 56.202 & 79.730 & $56.184$ $(-0.018)$ & $79.716$ $(-0.014)$ & 71 & 290 \pm 7 & 197 & 807 \pm 19  & 107.5 & 17.9 & 161.0 \tabularnewline
0.35-192 & 58.582 & 81.252 & $58.586$ $(+0.004)$ & $81.252$ $(-0.000)$ & 103 & 422 \pm 23 & 278 & 1141 \pm 22  & 152.2 & 24.2 & 239.2 \tabularnewline
0.35-224 & 60.384 & 82.750 & $60.394$ $(+0.010)$ & $82.768$ $(+0.018)$ & 129 & 529 \pm 18 & 381 & 1559 \pm 27  & 206.9 & 56.9 & 324.3 \tabularnewline

		\bottomrule
	\end{tabular}
\end{table*}
\fi

%% file: extensions.tex
\section{Conclusion and Future Work} Homomorphic encryption is a promising
solution to preserving privacy of user data during DL inference.
Current DL solutions using HE induce significant slowdown and memory overhead
compared to performing inference on unencrypted data. One potential solution to
this overhead is the use of plaintext packing, which enables storing multiple
scalars in a single plaintext or ciphertext. The choice of how to use plaintext
packing typically either increases throughput, via batch-axis plaintext packing,
or reduces latency, via inter-axis plaintext packing.

In this work, we presented \ShortTitle, which introduced several optimizations
to SEAL's implementation of the CKKS encryption scheme, for batch-axis plaintext
packing. Our optimizations result in a \tocheck{3x-88x} improvement in scalar
encoding, a \tocheck{2.6x-4.2x} speedup in ciphertext-plaintext scalar addition,
and a \tocheck{2.6x} speedup in ciphertext-plaintext multiplication.

We also introduced lazy rescaling, a CKKS-specific graph-based optimization
which reduces the latency by \tocheck{8x} on the CryptoNets network.
Additionally, we introduced complex packing, which doubles the throughput with
minimal effect on runtime.

Together, these optimizations enable state-of-the art throughput of
{\SI[allow-number-unit-breaks]{1998}\imagespersec} for the CryptoNets network
for the MNIST dataset. Furthermore, the integration of our approach with
nGraph-HE enables inference on pre-trained DL models without modification. To
demonstrate this capability, we presented the first evaluation of MobileNetV2,
the largest DL model with linear layers evaluated homomorphically, with
\tocheck{60.4}\percent/\tocheck{82.7}\percent\ top-1/top-5 accuracy, and
amortized runtime of \SI{381}\msperimage. To our knowledge, this is also the
first evaluation of a model with encrypted ImageNet data.

One avenue for future work involves performing non-polynomial activations
securely. In our approach, a client computes activations such as MaxPool and ReLU by first decrypting, the computing the non-linearity in plaintext, then encrypting the result. In the near future, we
plan to add support for other privacy-preserving primitives, e.g., Yao's Garbled
Circuit, to provide a provably privacy-preserving solution. Other
directions for future work include further optimization of scalar encoding for
complex numbers, and optimizing plaintext-ciphertext addition and multiplication
with Intel Advanced Vector Extensions (AVX).


%% file: appendix.tex
\section{Appendix}
\subsection{Network Architectures}
\label{app:architectures}
For each architecture, $n$ indicates the batch size.
\begin{itemize}
	\item
	\label{arch:cryptonets}
	CryptoNets, with activation $Act(x) = x^2$.
	\begin{enumerate}
		\item \emph{Conv}. [Input: $n \times 28 \times 28$; stride: 2; window: $5
		\times 5$; filters: 5, output: $n \times 845$] + \emph{Act}.
		\item \emph{FC}. [Input: $n \times 845$; output: $n \times 100$] + \emph{
		Act}.
		\item \emph{FC}. [Input: $n \times 100$; output: $n \times 10$].
	\end{enumerate}

	\item
	\label{arch:cryptonets-relu}
	CryptoNets-ReLU, with activation $Act(x) = ReLU(x)$.
	\begin{enumerate}
		\item \emph{Conv with bias}. [Input: $n \times 28 \times 28$; stride: 2;
		window: $5 \times 5$; filters: 5, output: $n \times 845$] + \emph{Act}.
		\item \emph{FC with bias}. [Input: $n \times 845$; output: $n \times 100$] +
		\emph{ Act}.
		\item \emph{FC with bias}. [Input: $n \times 100$; output: $n \times 10$].
	\end{enumerate}
\end{itemize}

\subsection{Parallel Scaling}
\label{app:parallel_scaling}
\begin{figure}[h]
	\begin{center}
		\includegraphics[width=\columnwidth]{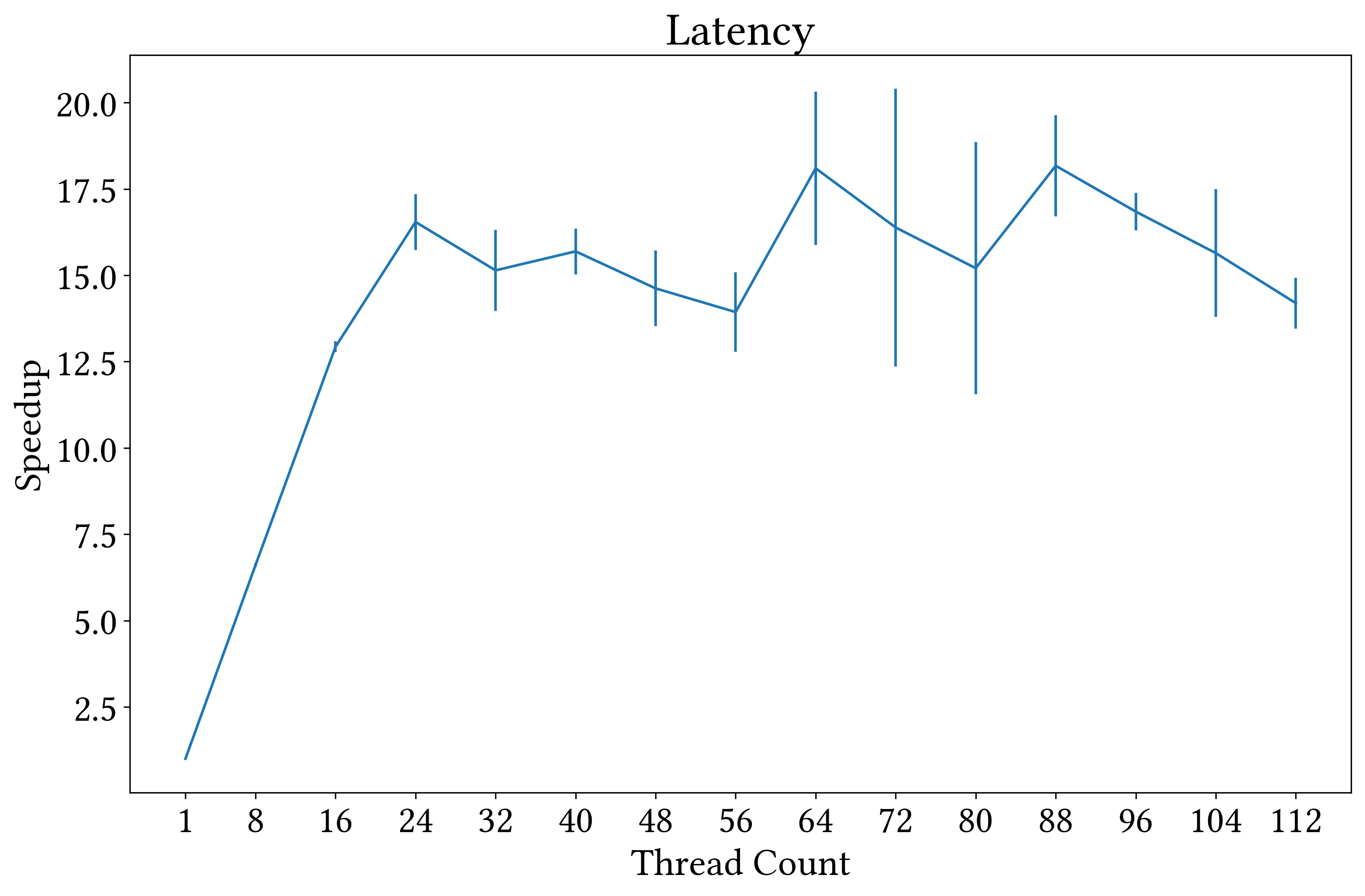}
		\caption{Runtimes on CryptoNets network with different number of threads. Runtimes are averaged across 10 trials.}
		\label{fig:cryptonets_threads_scaling}
	\end{center}
\end{figure}

nGraph-HE~\cite{boemer2019ngraph} uses OpenMP~\cite{dagum1998openmp} to
parallelize high-level operations such as Dot and Convolution. As such, the
runtime depends heavily on the number of threads. For the CryptoNets network
with $N=2^{13}, L=6$, 
Figure~\ref{fig:cryptonets_threads_scaling} shows the latency decreases linearly
with the number of threads up to thread count 16. Best performance is achieved
with 88 threads. However, the performance with 24 threads is just 9\percent\ 
slower (1.87\Sec\ vs. 2.05\Sec) than with 88 threads, representing a better runtime-resource tradeoff. In general, the optimal number of threads will depend on the network.

\subsection{Scalar Encoding}
\label{app:scalar_encoding_proof}

\begin{lemma}
Refer to Algorithm~\ref{alg:encode_plain_general} for the general CKKS encoding
algorithm. If the input vector $c$ consists of the same real number $r$ in each
slot, then the output plaintext $p$ will contain the same real number in each
slot.
\begin{proof}
	We refer to the notation in Algorithm~\ref{alg:encode_plain_general}. Since $c
		\in \mathbb{R}^{N/2}$, $c=c^*$, and so line~\ref{line:gen_encode:c} and
		line~\ref{line:gen_encode:c_conj} yield $p \gets (r, r, \ldots, r)$, the
		same value in every slot. Now, we show
		$$\text{DFT}^{-1}(r, r ,\ldots, r) = (r, 0,
		0, \ldots, 0).$$
		The DFT$^{-1}$ can be represented by a matrix
		transformation $W \in \mathbb{C}^{N \times N}$ with $W = \big(w_{jk}
		\big)_{0\leq j,k \leq N-1}$ for $w_{jk} = \frac{\omega^{-jk}}{N}$ where
		$\omega=e^{-2\pi i/N}$ is a primitive $N^{\text{th}}$ root of unity. In
		particular, the first row of $W$ consists of all ones, and the sum of every
		$j^{\text{th}}$ row for $j \neq 0$ is 0, since
		\begin{align*}
		\sum_{k=0}^{N-1} \frac{\omega^{-jk}}{N} &= \frac{1}{N} \left( \frac{\omega^j(1-\omega^{-jN})}{\omega^j - 1}\right) \\
		&= 0
		\end{align*}
		where the last equality uses that $\omega$ is a root of unity. Now, since
		$p$ has all the same values,
		$$\text{DFT}^{-1}\left(r, \ldots, r \right) = \big( \sum_i p_i / N,
		0, \ldots, 0 \big) = (r, 0, \ldots, 0).$$
		Scaling by $s$ yields $( rs, 0, \ldots, 0)$.
		The modulus reduction (line \ref{line:gen_encode:mod}) yields $([rs]_q, 0,
		\ldots, 0)$. Finally, the negacyclic NTT (line \ref{line:negacyclicntt}) can
		also be represented by a matrix transformation in the finite field
		$\mathbb{Z}/q\ZZ$, the integers modulo $q$. As with the DFT$^{-1}$ matrix,
		the first row is all ones, hence $$\text{NegacyclicNTT}(([ rs]_q, 0, \ldots, 0)) = [rs]_q (1,
		1, \ldots, 1).$$ Thus, the CKKS encoding has the same scalar, $[ rs]_q$, in
		each slot.
\end{proof}
\end{lemma}

\subsection{SEAL Performance Test}
\label{app:seal_perf_test}
Table~\ref{tab:seal_perf_test} shows the runtimes from SEAL's CKKS performance tests. The runtime increases with $N$ and $L$. In general, larger $L$ supports more multiplications. However, to maintain the same security level, $N$ must be increased accordingly.

\begin{table}[h!]
	\begin{center}
			\caption{SEAL CKKS performance test. Parameters satisfy $\lambda=128$-bit security. Runtimes averaged across 1000 trials.}
			\label{tab:seal_perf_test}
			\begin{tabular}{l 
					S[separate-uncertainty,
					table-number-alignment = center,
					table-figures-integer = 4,
					table-figures-decimal = 0,
					table-figures-uncertainty = 0]
					 S[separate-uncertainty,
					 table-number-alignment = center,
					 table-figures-integer = 5,
					 table-figures-decimal = 0,
					 table-figures-uncertainty = 0]
					  S[separate-uncertainty,
					  table-number-alignment = center,
					  table-figures-integer = 6,
					  table-figures-decimal = 0,
					  table-figures-uncertainty = 0]}
				\toprule
				 \centering \multirow{3}{*}{\textbf{Operation}} & \multicolumn{3}{c}{\textbf{Runtime
				 ($\mics$)}} \\
				 \cmidrule{2-4}
				& {\multirow{2}{*}{\thead{$N=2^{12}$ \\ $L=2^{  }$}}} &
				{\multirow{2}{*}{\thead{$N=2^{13}$ \\ $L=4^{  }$}}} &
				{\multirow{2}{*}{\thead{$N=2^{14}$ \\ $L=8^{  }$}}} \\ \\
				 \midrule
				 Add                & 16 & 59 & 237 \\
				 Multiply plain  & 58 & 234 & 936 \\
				 Decrypt          & 54 & 214 & 883 \\
				 Square           & 105 & 476 & 2411 \\
				 Multiply          & 155 & 709 & 3482 \\
				 Rescale           & 440 & 2224 & 10189 \\
				 Encode           & 1654 & 4029 & 10989 \\
				 Encrypt          & 1514 & 4808 & 16941 \\
				 Relinearize     & 936 & 4636 & 27681 \\
				 Decode           & 2153 & 7372 & 30175 \\
				 Rotate one step & 1098 & 5294 & 30338 \\
				 Rotate random & 4683 & 25270 & 158905 \\
				\bottomrule
			\end{tabular}
	\end{center}
\end{table}